\documentclass[aps,pre,twocolumn,groupedaddress,showpacs]{revtex4}

\usepackage{graphicx}
\usepackage{epsfig}
\usepackage{amsmath, amssymb}
\numberwithin{equation}{section} %add sections to equation numbers, %e.g. (IV.18)

\begin{document}

\newcommand {\be}{\begin{equation}}
\newcommand {\ee} {\end{equation}}
\newcommand {\bea}{\begin{eqnarray}}
\newcommand {\eea} {\end{eqnarray}}
\newcommand {\Eq}[1] {Eq.~(\ref{#1})}
\newcommand {\Fig}[1] {Fig.~\ref{#1}}
\def \bp {\mbox{\boldmath $\partial$}}
\def \q {{\bf q}}
\def \k {{\bf k}}
\def \zero {{\bf 0}}
\def \bfw {{\bf w}}
\def \bfx {{\bf x}}
\def \bfy {{\bf y}}
\def \bfz {{\bf z}}
\def \rb {\rm b}
\def \rg {\rm g}

\def \p {\ell}
\def \pt {p_0}
\def \pc {p_c}
\def \bt {{\bf T}}
\def \ct {{\mathcal T}}

\def \bigO {\mathcal{O}}
\def \indic {\mathbb{I}}
\def \calC {\mathcal{C}}
\def \calM {\mathcal{M}}
\def \ta {\widetilde{a}}
\def \tb {\widetilde{b}}
\def \tPhi {\widetilde{\Phi}}
\def \cE {\mathcal{E}}
\def \cG {\mathcal{G}}
\def \cT {\mathcal{T}}

%\ifpdf
%\DeclareGraphicsExtensions{.pdf, .jpg, .tif}
%\else
\DeclareGraphicsExtensions{.eps, .jpg}
%\fi

\title{Theory of minimum spanning trees I: Mean-field theory and strongly disordered spin-glass model}

\author{T. S. Jackson} \email[]{thomas.s.jackson@yale.edu}
\author{N. Read}
\email[]{nicholas.read@yale.edu} \affiliation{Department of
Physics, Yale University, P.O. Box 208120, New Haven, CT
06520-8120, USA}
\date{\today}

\begin{abstract}
The minimum spanning tree (MST) is a combinatorial optimization problem: given a connected graph with a real weight (``cost'') on each edge, find the spanning tree that minimizes the sum of the total cost of the occupied edges. We consider the random MST, in which the edge costs are (quenched) independent random variables. There is a strongly-disordered spin-glass model due to Newman and Stein [Phys.\ Rev.\ Lett.\ {\bf 72}, 2286 (1994)],
which maps precisely onto the random MST. We study scaling properties of random MSTs using a relation between Kruskal's greedy algorithm for finding the MST, and bond percolation. We solve the random MST problem on the Bethe lattice (BL) with appropriate wired boundary conditions and calculate the fractal dimension $D=6$ of the connected components. Viewed as a mean-field theory, the result implies that on a lattice in Euclidean space of dimension $d$, there are of order $W^{d-D}$ large connected components of the random MST inside a window of size $W$, and that $d = d_c = D = 6$ is a critical dimension. This differs from the value $8$ suggested by Newman and Stein. We also critique the original argument for $8$, and provide an improved scaling argument that again yields $d_c=6$. The result implies that the strongly-disordered spin-glass model has many ground states for $d>6$, and only of order one below six. The results for MSTs also apply on the Poisson-weighted infinite tree, which is a mean-field approach to the continuum model of MSTs in Euclidean space, and is a limit of the BL. In a companion paper we develop an $\varepsilon=6-d$ expansion for the random MST on critical percolation clusters.
\end{abstract}

\maketitle

\section{Introduction}
\label{s_intro}

\subsection{Motivation and approach}

The minimum spanning tree (MST) problem is one of the oldest and best-studied problems of combinatorial optimization \cite{lawler,papst,tarj,ccps,steele} and has found application to the physics of random systems \cite{cieplak_1,cieplak_2,ns_1,ns_2}. To define the problem, we consider an undirected, connected graph $G$ with vertex set $V$, edge set $E$ and a real-valued cost $\ell_e$ assigned to each edge $e \in E$. A spanning tree is then defined as a subset of the edges of $G$ that connects all the vertices and contains no cycles: in other words, it is a tree and it spans $V$. Such a tree must exist because the graph is assumed connected. A \emph{minimum} spanning tree $\bt$ is a spanning tree such that the sum of the costs of its edges,
\be
\label{eq_costfn} \ell(\bt) = \sum_{e
\in \bt} \ell_e, \ee
is minimized over the set of all spanning trees on G. If the costs $\ell_e$ are strictly positive, then any spanning subset of the edges that has minimum cost is automatically a tree.

If we view the cost \eqref{eq_costfn} as an energy, then this is a problem of finding the ground state of a classical system in which the configurations are spanning trees. If the costs of the edges are taken to be random variables, then this becomes a classical system with quenched disorder: the minimum must be found for a fixed set of costs, before any averaging over realizations of the edge costs is performed.

%referee
%three following paragraphs replace the one commented out afterwards

In the present paper and its companion \cite{jrII}, we begin a program to develop an analytical theory of the statistical geometry of random MSTs on a lattice $\mathbb Z^d$ in $d$-dimensional Euclidean space, where the edge costs are independently and identically distributed (iid) with a continuous probability distribution. (For continuous distributions of iid edge costs, the MST on a finite graph is unique with probability one.) We will primarily be interested in this model's long-range scaling properties (to be defined in the following section) because these are predicted to be the same for all models in the same universality class: in a quantitative sense, changing microscopic details of the model such as the type of lattice will not affect these properties (see e.g.\ Ref.\ \cite{scaling} for further discussion).
We will argue below that, due to universality, our results also apply to related problems, such as the continuum model \cite{as,steele}. In this model, the vertices are points in Euclidean space $\mathbb R^d$ which are Poisson distributed with uniform density, so the graph is the infinite complete graph, and the cost assigned to each edge is the Euclidean distance between its endpoints. Results concerning the expectation value of the cost $\ell(\bt)$ were previously given in Ref.\ \cite{read1}.
%end addition

%In the present paper and its companion (in preparation), we begin a program to develop an analytical theory of the statistical geometry of random MSTs on a lattice $\mathbb Z^d$ in $d$-dimensional Euclidean space, where the edge costs are independently and identically distributed (iid) with a continuous probability distribution. (For continuous distributions of iid edge costs, the MST on a finite graph is unique with probability one.) We will argue below that, due to universality, our results also apply to related problems, such as the continuum model \cite{as,steele}. In this model, the vertices are points in Euclidean space $\mathbb R^d$ which are Poisson distributed with uniform density, so the graph is the infinite complete graph, and the cost assigned to each edge is the Euclidean distance between its endpoints. Results concerning the expectation value of the cost $\ell(\bt)$ were previously given in Ref.\ \cite{read1}.

It is simple to solve computationally an instance of the MST problem, and many efficient algorithms exist \cite{krus,prim_1,prim_2,prim_3,tarj}. However, we are interested in the statistical properties of the random MST, which is relevant to physical disordered systems. MSTs play a role in transport in disordered networks \cite{bara,middle,optpath_1,optpath_2,optpath_3}, for example in current flow in random resistor networks \cite{wu_superhwy_1,wu_superhwy_2,read1,wu_rrns}. There exist in the literature several numerical simulations \cite{cieplak_1,cieplak_2,dd,ww_1,ww_2} determining $D_p$, the fractal dimension of the optimal path.

Furthermore, Newman and Stein (hereafter referred to as NS) have shown \cite{ns_1,ns_2} that in the strong disorder limit, the problem of finding the ground state of an Ising spin glass may be directly mapped on to the MST problem. Some of the results presented here bear directly on the discussion by these authors, which relate to fundamental questions about the ground state structure of spin glasses, and, by extension, solutions to hard optimization problems \cite{fu_and,mpv,nsrev}. Related considerations also arose in the quantum spin glass (or random transverse field Ising model) \cite{mmhf}. These connections will be discussed at length below.

Our program for MSTs takes a form similar to those for critical phenomena in, for example, Ising spin systems, or percolation. It proceeds in stages. The first stage is a mean-field theory, which we develop in the present paper by introducing and solving a version of random MST on the Bethe lattice (BL) (or Cayley tree), with suitable boundary conditions. The BL model can be solved exactly in some cases, or exactly for asymptotic, universal properties in more cases. We also investigate the related Poisson-weighted infinite tree (PWIT) model \cite{ald1,ald2,as}, which can be obtained as a limit of the BL as the coordination number goes to infinity and which can be viewed as the mean-field theory of the continuum model defined above. The subsequent stages would be to develop a full statistical field theory (or at least, a perturbation expansion) for the geometric properties in Euclidean space (either for the lattice or continuum models). Then a renormalization group approach should first show that corrections to the mean-field theory do not change the universal asymptotic properties above some critical dimension $d_c$, and identify $d_c$. A third stage would be calculation of universal properties at $d<d_c$ in an epsilon expansion in powers of $\varepsilon=d_c-d$. A fourth stage would be to show that the epsilon expansion is Borel summable, so that it defines the true results. The second and later stages will not be completed in this paper.

In developing an approach to the MST problem using Kruskal's greedy algorithm \cite{krus}, which is related to bond percolation, we will be led to consider the process of finding the minimum spanning forest (MSF) on the clusters of bond percolation as a function of the probability $p$ that an edge is occupied. (A spanning \emph{forest} is a spanning, vertex-disjoint collection of trees.) We call this process MSF$(p)$ (note that MSF$(1)$ is the same as the MST). We will argue that the universal properties of MSF$(p)$ at any $p>p_c$, where $p_c$ is the percolation threshold of the model, are the same. At this time we have been able to develop the perturbation techniques as mentioned above only for MSF$(p)$ with $p\leq p_c$. The expansion at $p>p_c$ is more difficult. However, it has been argued that some properties for $p>p_c$ are the same as those for $p=p_c$ \cite{wu_superhwy_1,wu_superhwy_2,optpath_1,wu_rrns}. In the absence of a perturbation theory treatment, it is not clear if this is true, but certainly we find many indications on the BL that the vicinity of $p_c$ dominates the behavior in the region $p>p_c$. In any case, in the companion (to be referred to as II) to the present paper  we develop first a small-$p$ expansion that is exact for any $p$ on a finite graph, and then a perturbation expansion using modified Feynman diagram techniques valid for $p\leq p_c$. By using renormalization-group techniques, this yields an epsilon expansion for the fractal dimension of a path on MSF$(p_c)$.

\subsection{Outline and discussion of main results}

We now describe the main results of this work.
The BL is an infinite tree with fixed coordination number (degree) at each vertex. As it is itself a tree, the MST on such a graph would be simply the whole graph, so attention to the definition of boundary conditions in the spirit of \cite{ns_1,ns_2} is essential in producing nontrivial behavior. The boundary conditions can be introduced first on a finite version of the BL, and then the infinite size limit can be taken. Specifically, as discussed in detail in Sec.\ \ref{s_prelims}, we adopt a wired boundary condition, which has the result that instead of the MST, the minimum object is a spanning forest. (This MSF produced by the wired boundary condition should not be confused with that in the process MSF$(p)$ mentioned above; in that language, at present we are considering the minimum objects at $p=1$.) We are interested in the statistical geometry of this non-trivial random forest. As the size of the lattice goes to infinity, the statistical properties of the MSF have a well-defined limit. We call the number of vertices that are connected to the central site and lie within $m$ steps on the BL the ``mass'' $M(m)$ within $m$ steps. We can then calculate (among other things) its expectation value $\overline{M(m)}$, and we find that it scales as
\be
\overline{M(m)}\sim m^3
\ee
as $m\to\infty$. The same result holds for the PWIT.

Employing this result as a mean-field theory, the standard method (see, e.g., \cite{bl_to_euclid,ns_1,ns_2}) for transferring results from the BL to a Euclidean lattice entails that distance $m$ on the BL corresponds to distance squared, $m\sim R^2$ on the Euclidean lattice. We find that the expected mass within distance $R$ of the origin scales as
\be
\overline{M(R)}\sim R^6,
\ee
so that the tree has fractal dimension $D=6$. As the trees fill the lattice, this means that the expected number of connected components that intersect a ball of radius $R$, denoted $N(R)$, scales as
\be
N(R)\sim R^\#\sim R^{d-6},
\ee
as $R\to\infty$, so the tree ``proliferation exponent'' \cite{aizen_perc, stah} $\#=d-D=d-6$. Note that on the BL, $N(m)$ increases exponentially with $m$, but there is a power-law correction factor of $m^{-3}$ which produces the behavior relevant for the Euclidean lattice.

Two points should be explained here. One is that $N(R)$ may be
subject to boundary effects at the surface of the ball, so that
the number of components intersecting the ball is larger, maybe
$R^{d-1}$. Our result is expected to describe the number of \emph{large} components intersecting the ball, say those whose intersection with the ball is of linear size $R/2$
\cite{aizen_perc,stah}.

The second, very important, point is that the MST on a
finite connected portion of a Euclidean lattice is again a
connected object by definition (for conventional ``free'' boundary conditions). But as this portion becomes larger
and approaches the whole lattice, the path between any two
vertices on the tree may make larger and larger excursions, so
that in the limit, viewed locally on any finite length scale $R$, the
tree appears to be a forest of many connected components
\cite{ns_1,ns_2}. The use of the wired boundary condition is intended to simulate this possible effect, by producing such a forest in a finite system, though the properties near the wired boundary may differ from those near the boundary of a ball of radius $R$ inside a system of size much larger than $R$.

The proliferation exponent $\#$ cannot become negative, so even if the mean field theory is indeed valid in high dimensions, it must break down in sufficiently low dimensions. These notions parallel some in other problems of random fractal clusters, such as in percolation at threshold (the critical point). A non-zero proliferation exponent $\#$ is the geometric counterpart to the violation of hyperscaling relations in critical phenomena at $d>d_c$ \cite{aizen_perc,stah}---hyperscaling is obeyed when $\#=0$. In critical percolation, the clusters at threshold are non--space-filling, and their fractal dimension is $D=4$ for $d\geq d_c$, while $d_c=6$. The MST, on the other hand, is an example of the subclass of such problems in which the union of the clusters fills space, so that $D+\#=d$. This then suggests that $d_c =D=6$ is the upper critical dimension, below which exponents (such as $D$) must change with $d$. Below $d=6$ it is plausible that there is only a single connected component (in the local sense described above) with probability one, or at least that, as $R\to\infty$, $N(R)<R^\alpha$ for any $\alpha>0$. Then $D=d$ and $\#=0$, and
\bea
D&=&\left\{\begin{array}{cl}6&\hbox{($d\geq 6$),}\\d&\hbox{($d\leq
6$);}\end{array}\right.\\
\#&=&\left\{\begin{array}{cl}d-6&\hbox{($d\geq
6$),}\\0&\hbox{($d\leq 6$).}\end{array}\right.
\eea

This result contrasts with that of NS, who suggested
that $d_c=8$ for MSTs \cite{ns_1,ns_2}, in the sense that $\#>0$ for
$d>8$, $\#=0$ for $d<8$. NS appear to have believed that the connected components of the MST have dimension $4$, but their arguments would actually imply that $D=8$ also; their arguments will be discussed in more depth in Sec.\ \ref{s_prelims}. (In fact, they showed that $d_c\leq 8$.) The question of the number of connected components was raised in a construction of a minimum spanning forest in the continuum model directly in infinite volume by Aldous and Steele \cite{as,steele}, who suggested that the forest has a single connected component in all dimensions $d$. It has been proved that there is a single connected component for this model in $d=2$ \cite{alex}, but for larger $d$ there has not so far been agreement even at a heuristic level. We note that for a different model of random trees on a lattice, that of uniform spanning trees (each spanning tree on a finite graph is given equal probability), similar behavior of the dimensions was proven, except that $d_{c\, \text{UST}}=4$ and $D_\text{UST}=4$ for $d>4$ \cite{pemantle}. Hence the universal properties of MSTs and
uniform spanning trees are distinct, at least in sufficiently high
dimensions \cite{ABNW,lps}.

In the companion paper II, we formulate a perturbation expansion for the geometry of the MSF$(p)$ process in Euclidean space for $p\leq p_c$. This is based on an exact small-$p$ expansion. It is used to calculate the fractal dimension of a path on the MST on a critical percolation cluster. Here we will mention only that our calculations can in principle be extended to other exponents, such as those defined in \cite{ABNW}, or carried to higher orders in $\varepsilon$. They can also be extended to include statistical properties that involve the cost of the MSF$(p_c)$. There do not appear to be any scaling relations that relate the geometric exponents for MSTs or MSF$(p_c)$ to those for percolation, unlike those found for the costs in \cite{read1}, even though the critical dimension $d_c=6$ is the same for all of them. In Ref.\ \cite{read1}, it was stated that there are different critical dimensions for properties involving the costs and for those involving only the geometry ($6$ and $8$, respectively). This remark was based on NS's results for the geometry, and is superseded by the present paper.

\subsection{Structure of the paper}
\label{s_summ_results}
The remainder of the paper parallels the discussion of the preceding section. In Section \ref{s_prelims}, we discuss the general properties of MSTs that we exploit in our calculations, and explain the relation
to percolation. We introduce Prim's and Kruskal's algorithms for computing the MST; these are related to invasion percolation and ordinary bond percolation, respectively. Similarly, the continuum model of MST is related to continuum versions of these. The strongly-disordered spin-glass model of NS is defined, and we provide a critique of their results for MSTs. We also give a scaling argument that suggests that the correct answer is $d_c=6$, $D=6$, which is confirmed in the following section. In Section \ref{s_bethe}, we solve the MST problem statistically on the BL using the connection between Kruskal's algorithm and percolation; this defines a mean field
theory. We also discuss the limit that gives MSTs on
PWITs. Implications of our results for the
strongly-disordered spin glass model are discussed in the
Conclusion.

%%%%%%%%%%%%%%%%%%%%%%%%%%%%%%%%%%%%%%%%%%%%%%%%

\section{Minimum Spanning Trees, Percolation, and Strongly-Disordered
Spin Glasses}
\label{s_prelims}

In this Section, we describe basic properties and techniques for
solving a MST problem. There are two main simple algorithms to
consider, Prim's and Kruskal's. Both are ``greedy'' in nature, and
when the edge costs are assumed to be iid random variables with a
continuous distribution, both are related \cite{hartigan} to
models of percolation. Prim's algorithm, which seems to be more
popular in the physics literature on MSTs, is connected with
invasion percolation. Kruskal's algorithm (also sometimes referred
to simply as the greedy algorithm) is related to bond percolation.
As the latter is much more tractable from the statistical
mechanics point of view, this is the approach that we will use.
Here we review these connections and basic properties of
percolation. We review the strongly-disordered spin glass model of
Newman and Stein (NS), and critique their argument that the
critical dimension of the model is $d_c=8$. In NS, this critical
dimension is that dimension $d$ above which there is an infinite
number of ground states in the thermodynamic limit of the model.
We give a scaling argument for $d_c=6$.

\subsection{Basic properties and algorithms}
\label{s_mst_basic}

A property that is useful in finding the MST of an edge-weighted finite graph is the following \cite{papst}: Suppose a spanning forest is
given, that is not necessarily minimum cost (some connected
components of such a forest may consist of a single vertex and no
edges). If we choose a connected component, and then find the
edge $e$ of minimum cost among all those edges with just one end
in that component, then among all the spanning trees that contain
the given spanning forest (as a subset of the edges of the tree),
the one of minimum cost contains the edge $e$. Thus starting from
any given forest, one can greedily add edges of minimum cost (among those
that leave any one connected component at each stage) and arrive
at the spanning tree that has minimum cost among all those containing the given forest. In particular, starting with the forest in which each tree is a single vertex and no edges, one can find the MST. This still
leaves many ways to proceed in selecting the connected component
at each step. Two particular ways of doing so are of interest
here: a) the Kruskal or greedy algorithm \cite{krus}, in which at
each step one adds the cheapest edge not yet occupied that does
not create a cycle when added to the set of occupied edges; b) the
Jarn\'\i k-Prim-Dijkstra algorithm \cite{prim_1,prim_2,prim_3} (usually referred to as Prim's), in which one starts by selecting a vertex, and then adds the cheapest edge leaving the connected component containing
the initial vertex at each step. In either case, the algorithm
stops when a spanning tree is formed, that is when $|V|-1$ edges
are occupied. From the preceding result, both of these ultimately
produce the MST. (We have neglected to specify how to break any
ties that arise from edges of equal cost, which can lead to
non-unique MSTs, as these are not of interest in this paper.) The
difference between the two in terms of the geometry of the
connected components or ``clusters'' is that in Kruskal's
algorithm, at a typical stage there are several trees containing
more than one vertex, whereas in Prim's there is always only one.
Both procedures have similar running times, and can also be
improved \cite{tarj}.

The graph and the set of edge costs defines the MST via minimization of the total cost, which is the sum of the costs of the occupied edges as
in eq.\ \eqref{eq_costfn}. From the above algorithms, it is clear that the precise values of the costs are not important in finding the geometry of the MST (and hence neither is the probability distribution for them \cite{dd}). Only the rank ordering of the set of costs is important. Indeed, even this much information is not required,
as evidenced by the fact that the most efficient known
deterministic algorithm for computing the MST on an arbitrary
graph \cite{chazelle} has significantly faster asymptotic running
time than that for sorting all the edges of that graph by cost.

\subsection{Relation of Kruskal's algorithm with bond percolation}
\label{s_krus_vs_perc}

The standard way to define bond (or Bernoulli) percolation on a
graph is to declare that edges are either occupied with probability $p$ or not occupied,
independently. We are then
interested in the geometric properties of the clusters (connected components) formed by the occupied edges. For a graph
that is a portion of a lattice in Euclidean space, the basic
problem is to find the {\em percolation threshold}, the value of
$p$ above which there exists a path on a single cluster from one
side of the system to the other. When the graph is the complete
graph, the percolation problem becomes the theory of ``random
graphs'' (these graphs being the clusters). The threshold is then
the value of $p$ at which a cluster of size of order $|V|$ occurs,
as $|V|\to\infty$. In both cases, the behavior near the threshold
is a critical phenomenon, and there are critical exponents. The
complete graph model is one popular way of obtaining a mean-field
treatment of percolation.

A random edge-weighted graph gives rise to a sample of bond
percolation, as follows. We suppose that the weights or
costs on the edges are iid random variables, with a probability
density $P(\ell_e)$ of the cost of any one edge $e$. If we say that
all edges with costs less than or equal to $\ell_0$ are occupied,
then this gives a sample of bond percolation in which $p=p_0$ is
given by
\be%
p_0=\int_{-\infty}^{\ell_0} d\ell_e\, P(\ell_e). \label{p_0}\ee%
In particular, if $P(\ell_e)=1$ for $\ell_e$ in the interval
$[0,1]$ and zero outside, then $p_0=\ell_0$.

Now consider Kruskal's algorithm for such a random edge-weighted
graph. At each step, one must ``test'' the edges not tested
earlier, and find the cheapest one that does not create a cycle
when added to those occupied. We say that such an edge is
accepted. Thus we can view the algorithm as working through the
edges in order of increasing cost. If the algorithm is terminated
at cost $\ell_0$ (at which stage there may be only a spanning
forest, not yet a spanning tree), then the set of edges already
tested forms a sample of percolation, as just described. Thus, if
we had accepted all edges instead of only those that did not form
a cycle, we would have obtained a sample of bond percolation. We
can thus view percolation, as well as Kruskal's algorithm, as a
dynamical process in which clusters are grown beginning from the
set of vertices and no occupied edges, and eventually obtaining a
cluster spanning the graph. Kruskal's algorithm gives a variant on
this percolation process in which as $\ell_0$ increases, an edge
is accepted only if adding it does not form a cycle \cite{frieze}.
The process does not depend on the probability density
chosen for the costs, because as mentioned above only the ordering
is important for the tree. The process is fully characterized by
the variable $p_0$, and distinct probability densities $P(\ell_e)$
give rise to the same probability measure on MSTs, as emphasized
in Ref.\ \cite{dd}. Without loss of generality, we can consider
$\ell_e$ to be uniformly distributed between $0$ and $1$, so
$p_0=\ell_0$, as above. We will use the name MSF$(p_0)$ for the random spanning forest produced by halting the Kruskal algorithm at some value $p_0$; the true MST is obtained at $p_0=1$, or by halting when $|V|-1$ edges are occupied.

Later in the paper we develop analytical techniques to study MSTs
using the theory of bond percolation. Here we mention some of the
basic properties of percolation, for the limit of an infinite system \cite{stah}. The percolation threshold $p_c$ is non-universal; it depends on the lattice or class of graphs considered. For hypercubic graphs in dimension $d$ (i.e.\ ${\mathbb Z}^d$, with edges connecting nearest neighbors only), $p_c$ lies strictly between $0$ and $1$ for
$d>1$, while for $d=1$, $p_c=1$. For $p<p_c$, there are finite
clusters only. The typical size $\xi$ of a cluster (the
correlation length) diverges as $\xi\sim (p_c-p)^{-\nu_{\rm
perc}}$ as $p\to p_c$ from below. For $p>p_c$ (so assuming $d>1$),
there is a single infinite cluster with probability one, and for
$p_c<p<1$ a non-zero density of finite clusters (clusters not
connected to infinity). The typical size of the finite clusters
now defines $\xi$, and $\xi\sim(p-p_c)^{-\nu_\text{perc}}$ as $p\to
p_c$ from above, with the same exponent $\nu_\text{perc}$. The
correlation length exponent $\nu_\text{perc}$ equals $1/2$ for
$d\geq 6$, but deviates from this value for $d<6$. At criticality,
$p=p_c$, there is a power law distribution of cluster sizes up to
infinite size. For $d\geq6$ these clusters have fractal dimension
$D_\text{perc}=4$, while this dimension $D_\text{perc}$ deviates
from this value below $d=6$. For $d\geq 6$, there are of order
$R^{\#_\text{perc}}$ large such clusters intersecting a ball of
radius $R$, where $\#_\text{perc}=d-6$, while $\#_\text{perc}=0$ for
$d\leq 6$. All the exponents may be calculated by considering
suitable correlation functions. (All power laws may be subject to
logarithmic corrections when $d=6$, which we do not consider.)

\subsection{Relation of Prim's algorithm with invasion
percolation}

The arguments of the previous subsection can be repeated for
Prim's algorithm. From Prim's algorithm, one obtains a dynamical
process in which at each step a single edge is added to growing
cluster (tree), which contains the initial vertex $\bfx$. The edge
added is the cheapest one that borders the current tree and does
not form a cycle. In a finite system, this cluster eventually
spans all the vertices, and at that point becomes the MST. In an
infinite system, if the ``time'' in the process is identified with
the number of edges added, then the tree continues to grow
indefinitely until an infinite tree is obtained after infinite
time. We should note that it is not obvious that this infinite
tree is spanning, and for $d > 1$ it will not be. Because
the tree depends on the starting vertex $\bf x$ say, we denote
this tree obtained after infinite time as $T_\infty({\bf x})$; it
is a random object that depends on the costs of all the edges that
were tested in growing the tree. The latter edges are those on the
tree, together with those that have at least one end connected
to $\bfx$ by $T_\infty(\bfx)$.

Prim's algorithm is connected to a type of percolation called
invasion percolation \cite{inv_1,inv_2,inv_3} analogously to how Kruskal's
is connected to bond percolation; for Prim's algorithm this
connection has been noted repeatedly in the physics literature
\cite{ns_1,ns_2,cieplak_1,cieplak_2,bara,dd}.
Using the edge-weighted graph with iid costs as before, a model of
invasion percolation is obtained by modifying Prim's algorithm to
accept the cheapest edge that borders but is not on the current
cluster (i.e.\ by neglecting the no-cycle condition). Notice that
in both the Prim and invasion process, the costs of the accepted
edges do not increase monotonically as they did in the case of the
Kruskal/percolation process, because when the cluster first enters
a region of space, additional edges become available, which may be
cheaper than those accepted earlier. Indeed, in an infinite
system, the edges accepted after long times have (with high probability) costs $\leq p_c$ of the corresponding bond percolation problem (in the model with costs uniformly distributed between $0$ and $1$) \cite{inv_1,inv_2,inv_3,ccn}. It is believed \cite{inv_1,inv_2,inv_3,ccn} that the invasion cluster is a fractal very similar to the critical percolation clusters, with the same universal properties, so its fractal dimension is $D_\text{inv}=D_\text{perc} =4$ for $d\geq 6$. Clearly, the set of vertices on the invasion cluster and on the invasion tree produced by
Prim's algorithm are the same, so the same fractal dimensions apply to the trees $T_\infty(\bfx)$.

\subsection{Strongly-disordered model of a spin glass}
\label{s_intro_sdsg}

NS \cite{ns_1,ns_2} (see also Ref.\ \cite{cieplak_1,cieplak_2}) defined a strongly-disordered spin glass model and showed that the problem of finding the ground state maps onto a MST problem. We now briefly describe
this model. In the next subsection, we discuss the arguments of
NS, which motivated many of the questions addressed in this paper.

The Edwards-Anderson (EA) Ising spin glass model \cite{mpv} is defined by the
Hamiltonian%
\be%
{\cal H}=-\sum_{i<j}J_{ij}s_is_j\ee%
for Ising spins $s_i=\pm 1$, where $J_{ij}$ are quenched random variables.
The positions $i$, $j$ are taken as lattice points in a portion
(say, a cube of side $L$) of a $d$-dimensional hypercubic lattice,
and for the edges $\langle ij\rangle$ of the corresponding graph
(i.e.\ ``nearest neighbor bonds''), the $J_{ij}$'s are iid
variables with a distribution independent of the portion of the
lattice chosen (in particular, independent of the number $|V|=L^d$
of spins), for example a Gaussian distribution with mean zero and
standard deviation $J_0$. The $J_{ij}$'s are quenched random
variables, meaning that thermodynamic quantities must be
calculated with a fixed sample of $J_{ij}$'s, and then averages (or
moments, etc) taken at the end. The NS strongly-disordered model
differs from the EA model in the distribution of $J_{ij}$s. While
they are still iid, the width of the distribution is assumed to be
extremely large, and depends on the number of spins, in a fashion
to be specified below. The distribution is symmetric, so that the
sign of $J_{ij}$, $\epsilon_{ij}\equiv{\rm sgn}\,J_{ij}$, is $\pm
1$ with probability $1/2$ for either case, independently of the
random magnitude $K_{ij}=|J_{ij}|$. NS focus on ground state
properties, and specify a boundary condition that the spins on the
boundary are an arbitrary set of values $\pm 1$, chosen
independently of the $J_{ij}$s on the edges in the interior.
Clearly, similar models can be defined on other graphs, including
the complete graph (infinite-range model), or with different
boundary conditions.

The central idea in the use of a broad distribution of disorder
(that is, of the $K_{ij}$'s) is that for such a broad distribution,
for any subset of edges, there is always a single $K_{ij}$ that
dominates all others in the set, or even dominates the sum of all
the others. Thus the width of the distribution must simply be chosen large enough that this is so, and that is why the width must increase
with the system size. In this case the problem of finding the
ground state of the model with given bonds $J_{ij}$ and boundary
spin values may be solved by a greedy procedure. NS chose to use
a procedure similar to Prim's algorithm. To find the orientation
(i.e.\ the value $\pm 1$) in the ground state of a given spin
located at $\bfx$, first find the largest $K_{ij}$ among the edges
leaving $\bfx$. The relative orientation, $s_is_j$ for $i$
corresponding to $\bfx$, is clearly then determined, but not $s_i$
itself. Then find the largest $K_{ij}$ leaving this cluster, that
is with one end on the cluster, but not the other. This fixes the
relative orientation with a further spin $k$, say. This process
can be repeated, adding edges to the cluster until a boundary spin
is encountered, at which point all the spins on the cluster are
now determined. Edges that connect two sites of a cluster that are
already connected need not be considered (or accepted), since the
relative orientation of those spins is already determined. Thus
the edges accepted form a tree. The process is clearly identical
to Prim's algorithm, until the growing invasion tree touches the
boundary. The process can then be repeated starting with any spin
not already fixed, until all spins have been determined. For the
trees after the first, the process is defined to restart from
another vertex not already connected if the growing tree
encounters an earlier tree, as well as if it encounters the
boundary, since again such an encounter fixes the spin
orientations on that growing tree.

The process of repeatedly growing trees until every vertex is part
of a tree that touches the boundary exactly once produces a
spanning forest on the graph (or portion of the lattice). From the
percolation point of view, similar use of boundary conditions is
called a wired boundary condition, and has also been used in MST
problems \cite{steele}. We may imagine that the edges that connect
the boundary vertices to one another have costs less than all
those in the interior, so in Prim's algorithm, once the first tree
encounters the boundary, it is immediately connected to all other
boundary sites along the boundary. (The precise ordering of the
boundary costs among themselves is not of interest, because we are
not interested in the edges of the MST on the boundary.) Then as
all earlier trees are now viewed as connected in a single tree
that includes all the boundary vertices, the result when the
process terminates is a spanning tree. It is in fact the MST on
the given graph with costs $-K_{ij}$, together with costs
$\sim -\infty$ for the boundary edges. The process used by NS is
a variant on the Prim process that occasionally restarts from a
different vertex that is not connected to any others. This is
among the many different ways to find the same MST, as can be seen
using the general fact from Sec.\ \ref{s_mst_basic}. Hence {\em
any} valid construction of the MST with this wired boundary
condition produces the same spanning tree, and hence also produces
the same ground state of the NS model. In particular, we can use
the Kruskal algorithm.

Thus finding the ground state of the NS model is a MST problem.
Once the MST with the costs $-K_{ij}$ in the interior, and wired
boundary conditions, has been found, the spin orientations follow
using the signs $\epsilon_{ij}$ and the boundary spins. Because
this is now essentially solving a spin-glass model on a forest,
there is effectively no frustration left at this stage. We recall
that frustration in an Ising spin model with Hamiltonian of the
form of $\cal H$ means that there exist cycles of the graph such
that the product of signs $\epsilon_{ij}$ of the edges on the
cycle is negative; generally this means that all the details of
the magnitudes of the couplings must be studied in order to find
the ground state. For the EA model in more than $2$ dimensions,
finding the ground state is computationally costly. Indeed, in
$d>2$ the problem of determining whether the ground state energy
(cost) of an Ising spin glass Hamiltonian $\cal H$ is less than
some bound (budget) is NP-complete \cite{barahona}, and so
presumably cannot be solved in polynomial time (for a discussion
of NP-completeness, see Ref.\ \cite{bk_papa}). (For $d=2$, or for
planar graphs, the spin glass ground state can be mapped to a
network flow problem, and solved in polynomial time.) By contrast,
for the NS model, the ground state can be found in polynomial time
using an MST algorithm.

\subsection{Ground states of the NS model and fractal dimensions}
\label{s_ns_gs}

We now turn to NS's analysis of their model. The model was
constructed so as to be soluble. The use of fixed boundary spins,
or the wired boundary condition on the MST, was motivated by a
deep view of the meaning of a thermodynamic state in a spin
system, or in the present case a ground state \cite{nsrev,ns_1,ns_2}. In an infinite
system, a ground state can be defined as a spin configuration the
energy of which cannot be lowered by reversing the values of any
{\em finite} set of spins. A ground state spin configuration in
any bounded portion of the lattice (such a configuration is simply
that of minimum energy) is completely determined by the values of
the spins on its boundaries. As the size of this portion goes to infinity (keeping the bonds $J_{ij}$ the same in the interior as fresh
bonds and spins are added at the boundary), there should exist
sequences of boundary conditions such that the spin configuration
seen in any ``window'' (subregion of the system) converges to a
limit. When this is done for a sequence of windows diverging in
size (such that the spin configurations agree where the windows
overlap), then a ground state of the infinite system is obtained.
It is then clear that this ground state is determined by a choice
of boundary conditions infinitely far away. The use of arbitrary
boundary spins on the boundary of a finite region (a hypercube of
side $L$, say) as in NS is then part of this process, and
approximates the ground states of the thermodynamic limit.

As we have seen, NS determine the ground state of their model,
using boundary spin values and a set of invasion trees. Then the spins in the
box lie on a spanning forest in which each connected component
touches the boundary just once. All spins on such a connected
component will be reversed if the value of the corresponding
boundary spin is reversed. Thus the logarithm of the number of
possible distinct ground states obtained inside the box of size
$L$ by varying the boundary spin values is bounded by ${\cal O}(
L^{d-1}\ln 2)$ (it is a bound, not the actual number, because some
boundary spins may not be connected to any interior spins, and
these boundary spins are not to be considered when counting
configurations). But it is better to consider a window of side $W$
within the box, with $W\ll L$, and ask how many distinct
configurations can be obtained within the window as the boundary
conditions are varied, preferably without counting configurations
that differ only just inside the surface of the window. The
logarithm of this number is given by the number of connected
components of the minimum spanning forest that intersect the
window (neglecting those whose linear size is less than $W/2$,
say). (There may be some ambiguity here concerning connected
components that intersect the window more than once, and are
connected outside the window but not inside, as these do not give
$4$ ground states, but only $2$, however we will assume this is
not significant.) Thus this number $\ln {\cal N}(W)$ is the same
as $N(W)$ as defined in Sec.\ \ref{s_intro}:%
\be%
\ln {\cal N}(W)=N(W)\ln 2.\ee%
We have arrived at the same question about MSTs that was
already introduced in Sec.\ \ref{s_intro}: the behavior of the
number of connected components of a MST intersecting a window, or
alternatively the fractal dimension of the connected component(s)
of an infinite MST.

NS studied this question using the Prim algorithm. To find whether
two vertices, $\bfx$, $\bfy$, say, are on the same connected
component, we may grow the invasion tree from each of them, and
see whether they intersect. More exactly, we could first grow the
tree from $\bfx$ to infinity, then begin again from $\bfy$,
stopping if $T_\infty(\bfx)$ is encountered. Since each invasion
tree is a fractal of dimension $D_\text{inv}=4$ (for $d\geq6$), NS
were able to show rigorously that when $d>2D_\text{inv}=8$, there
is a non-zero probability that the two trees ``miss'' and never
intersect \cite{ns_1,ns_2}. This event implies that the two points are on
distinct connected components, and so if the critical dimension
$d_c$ is defined as that above which the MST has more than one
connected component in the thermodynamic limit (with probability
one), then NS proved the upper bound $d_c\leq 8$.

To determine the actual value of $d_c$, NS stated that they need a
converse result, in other words a lower bound. This converse
statement would result if the probability that the two invasion
trees do intersect is of order $1$ for $d<8$. Such behavior is
natural if one has two independently-grown fractals of dimension
$4$ (or more generally, if $2D>d$ for two fractals of dimension
$D$). Now for the invasion trees, independence does hold when the
clusters are sufficiently far apart, because the edges considered
when growing either one are only those bordering it, not all those in the system. (We recall that the edges bordering the tree are those
with one end on the tree, and one end off it.) More precisely,
they are independent as long as they do not intersect and the sets
of edges bordering one are disjoint from the set of edges
bordering the other. This leads to NS's result $d_c\leq 8$.
However, if the trees do become close together so that they share
one or more bordering edges, for example if $T_\infty(\bfx)$ is
already grown first, and we are growing the tree from $\bfy\not\in
T_\infty(\bfx)$, then they are correlated. From the invasion
process, we can see that edges bordering $T_\infty(\bfx)$ must be
of cost higher than $p_c$ (again, in the model where edge costs $\ell$ or $p$
are uniform between $0$ and $1$), and not arbitrary like those
bordering the tree from $\bfy$ when they are {\em first}
encountered. This very strong correlation effect reduces the
probability that any such edge that forms the connection between
$T_\infty(\bfx)$ and the tree from $\bfy$ will be accepted on
$T_\infty(\bfy)$. Hence we expect that $d_c<8$.

More quantitatively, as the tree $T_\infty(\bfx)$ has dimension
$D_{\rm inv}=4$ for $d>6$, the probability $G(\bfz-\bfx)$ that a point $\bfz$
will be on $T_\infty(\bfx)$ should behave as
\be
G(\bfz-\bfx) \sim |\bfz-\bfx|^{D_\text{inv}-d}
\ee
 for large $|\bfz-\bfx|$ \cite{stah,ns_1,ns_2}. If the two trees were grown independently, the
probability that they intersect would behave as
\be
\int d^d\bfz\,G(\bfz-\bfx)G(\bfz-\bfy)\sim |\bfx-\bfy|^{2D_\text{inv}-d}.
\ee
For $d<2D_\text{inv}(d)=8$ this is instead of order one.
This would give the converse result of NS. It would also imply
that connected components of the minimum spanning forest have
dimension at least $2D_\text{inv}=8$ for $d\geq 8$ (though NS seem
to have believed that this dimension would be $4$). But because
the trees are not independent when they intersect, neither result
is correct.

To make a crude estimate of the correct result, we use scaling
arguments for the growth of $T_\infty(\bfy)$. Given $\bfx$ and
$\bfy$, the tree from $\bfy$ first approaches close to
$T_\infty(\bfx)$ when its linear size $\xi$ is of order $|\bfx-\bfy|$.
The costs of edges typically accepted when it reaches a size scale
$\xi$ will be distributed up to about $p_c$, with $p-p_c$ not
larger than of order $\xi^{-1/\nu_\text{perc}}$, where we recall
that $\nu_\text{perc}=1/2$ for $d\geq 6$. (Here $p$ stands for the cost of the edge, in accordance with the earlier discussion of the relation of MST with percolation.) Indeed, the probability density for an edge to be accepted at this stage will be given by a scaling function of $(p-p_c)\xi^{-1/\nu_{\rm perc}}$, which goes to one when its argument is large and negative, and to zero when its argument is large and positive. Similarly, the costs of edges bordering $T_\infty(\bfx)$ at distance of order $\xi$ from $\bf x$ have a probability distribution that is a function of $(p-p_c)\xi^{-1/\nu_{\rm perc}}$, which vanishes for $p<p_c$ and goes to one for large positive arguments. Multiplying these functions and integrating over $p$ to obtain the probability that the first such edge is accepted, the result is of order $\xi^{-1/\nu_{\rm perc}}=|\bfx-\bfy|^{-2}$ for $d>6$.  Assuming that consideration of subsequent events, which lead to smaller probabilities as they occur on larger scales, leads to a converging sum of terms, this then leads to the reduction of the dimensions by $2$, so $d_c=6$ and $D=6$ for $d\geq 6$ (where, throughout this paper, $D$ is the dimension of the connected components of the MST). In the next
section, we obtain the same results by much more secure methods on
the BL, and a plausible application to Euclidean space
also.

The behavior of the probability that two vertices are on the same
connected component of the spanning forest is exactly what is
needed to count the significantly-different ground states of the
NS model that are visible in a window. The number of large
connected components intersecting the window will scale as
$W^{d-D}$, which by the above scaling argument, and the results
below, will be $W^{d-6}$ for $d\geq 6$; thus $\#=d-6$ (thus the use of $\nu_{\rm perc}(d)=1/2$ in the above scaling argument is justified self-consistently). We mention
here that the authors of Ref.\ \cite{dd}, who considered the path
exponent $D_p$ but not $D$ or $\#$ for MSTs, also stated that the
critical dimension for MSTs is $6$, apparently because they were
using Prim's algorithm, and its connection to invasion
percolation, for which again $d_c=6$.

%%%%%%%%%%%%%%%%%%%%%%%%%%%%%%%%%%%%%%%%%%%%%%%%%%%%%%%%%%%%%%%%

\section{Bethe lattice}
\label{s_bethe}

In this section we define and solve the MST problem on the BL with wired boundary conditions. The BL can be motivated by the desire to find a ``mean field'' theory, in which only the mean effect of neighbors of a vertex is included, while effects of correlations that propagate around cycles of the lattice are neglected. The BL fulfils these requirements as it possesses no cycles, but is still homogeneous due to the constant coordination number (degree). For the MST on the BL, we calculate the expectation value of the number of vertices that are connected to the origin within radius $m$. We show how to define and analyze other correlation functions also. The BL solution forms the basis for a mean-field theory defined on a finite-dimensional lattice in the next section. The BL results imply that the mean-field fractal dimension of paths on the MST in Euclidean space is $2$, and the fractal dimension of connected components is $6$. This establishes that $d_c = 6$ by the argument given above.

\subsection{Preliminaries}

The BL is a tree of degree $z=\sigma+1$ at every vertex, except for the ``leaves'' at the boundary, which are all a ``distance''
(distance means minimum number of steps along the tree required to
go from one vertex to another) $M$ from the central vertex, which
we call the origin. We use the term BL interchangeably
to refer to both a finite graph of radius $M$ and the graph
obtained in the $M \to \infty$ limit. Since the MST on a finite
tree graph will clearly be the whole tree, we must use wired
boundary conditions, as discussed in the previous section, in
order to obtain an interesting spanning forest. We will see that
the limit of such a forest does exist, at least in terms of its
local statistical properties, at finite distances from the origin
(hence, far inside the boundary); this is the ``weak limit'' which
has been discussed in Ref.\ \cite{lps}. We expect that the local
properties of this forest in the limit mimic the local properties
of the MST in Euclidean space in a mean-field sense.

\begin{figure}
\includegraphics[width=3in]{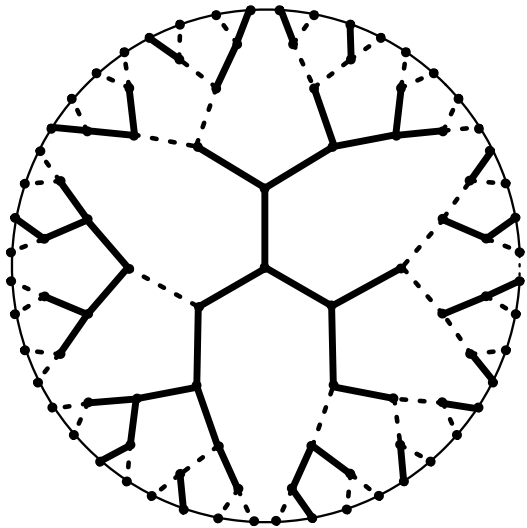}%
\caption{A sample realization of the MST (solid lines) on the
BL with wired boundary condition. The graph shown has
$\sigma = 2$ and radius $M = 5$. Vertices on the boundary are
connected with boundary edges of vanishing cost, shown here as
thin lines. All vertices in the interior are connected to the
boundary only once. \label{figbethe}}
\end{figure}

The wired boundary condition means that we connect all vertices on
the boundary of the BL to each other with additional
``boundary edges'' of vanishing cost, so that they are always
connected for $\pt >0$, as shown in figure \ref{figbethe}. In
practical terms, this means that when we run the Kruskal process,
once a connected component (or cluster) of the forest touches the
boundary, no edge that would connect it to the boundary by a
distinct path can be added later. When the process is run up to
$p_0=1$, we obtain a forest spanning the BL, which
becomes a tree if the boundary edges are included.

We will use the correspondence with bond
percolation discussed in section \ref{s_prelims}. Percolation on
the BL has been studied thoroughly \cite{fisher_essam,
stah}. The most basic question of interest is to find the
probability that a vertex in the interior is connected to the
boundary, as a function of the probability $p$ for each edge to be
(independently) occupied. A convenient quantity to look at is the
probability $F_M(p)$ that a given vertex is {\em not} connected to
the boundary at distance $M$ via a path of edges through a given
outward branch. This obeys the recurrence relation%
\be%
F_{M+1}(p)=1-p+pF_M(p)^\sigma,\ee%
with $F_0(p)=0$. This arises because either a) the first edge
along the branch is unoccupied, which occurs with probability
$1-p$, or b) the first edge is occupied (with probability $p$) and
there is no connection to the boundary through the $\sigma$
branches further towards the boundary; these two alternatives are
disjoint. As the radius $M$ of the lattice goes to $\infty$, all
vertices a fixed distance from the origin are far from the
boundary, and the probability of any of them not being connected
to the boundary along a particular outward branch approaches a
limit $\lim_{M\to\infty} F_M(p)=F(p)$, which is given by a {\em
stable} fixed point with $0\leq F\leq 1$ of the recurrence, that is
\be \label{eq_fdef} F(p) = 1- p +
pF(p)^{\sigma}. \ee %
This has the trivial solution $F=1$, and its stability is given by
linearizing the recurrence about this solution; the eigenvalue of
the linearized recurrence for $F_M-1$ is $\sigma p$. Thus for
$p<1/\sigma$ the solution $F=1$ is stable. In this regime, no
interior vertex is connected to infinity, with probability one.
The value $p_c=1/\sigma$ is called the percolation threshold. For
$p>p_c$, the solution $F=1$ is unstable, and the stable solution
is a non-trivial solution to the fixed-point condition, eq.\
(\ref{eq_fdef}). The existence, uniqueness, and stability of, and
the convergence to, these fixed points is proved in, for example,
Ref.\ \cite{arthney} (in different notation). For $\sigma=2$, this
solution can be found explicitly,%
\be%
F(p)=(1-p)/p\ee%
for $p>1/2$ ($\sigma=2$). For general $\sigma$, we can expand in
powers of $p-p_c$, and find that $1-F(p)=a(p-p_c)+{\cal
O}((p-p_c)^2)$ as $p\to p_c$ from above, where
$a=2\sigma/(\sigma-1)$. (As $p\to 1$, $F(p)\to0$.) The probability that any vertex
is connected to infinity (and hence is on an infinite spanning
cluster) is then $P_{\infty}(p) = 1-F^{\sigma+1}(p)$, which is
zero for $p<p_c$, and turns on with a discontinuous derivative at
$p_c$. This behavior defines the critical exponent $\beta=1$ via $P_\infty(p) \sim (p-p_c)^\beta$ for $p\to p_c$ from above, which illustrates that percolation is a critical phenomenon with
discontinuous properties at $p_c$.

%%%%%%%%%%%%%%%%%%%%%%%%%%%%%%%%%%%%%%%%%%%%%%%%%%%%%%%%%%%%%%

\subsection{Correlation functions}
\label{s_bethe_correl}

In using Kruskal's algorithm to calculate the MST on the BL, we need to keep track of when different vertices become
connected to the boundary as $p_0$ is raised from $0$ to $1$. As
an example, consider the probability $P_{(0;m)}(p_0)$ that the
path on the MST from the origin to the boundary has formed by time
$p_0$, that is when all edges of cost $\leq p_0$ have been tested,
and passes through a given vertex a distance $m$ from the
origin. (When we discuss such paths, we always mean a path that
does not backtrack on itself.) We may consider this as done in a
finite system, and we always assume that the limit $M\to\infty$ is
unproblematic. Indeed, we will see that the only ingredient
involving properties all the way out to the boundary is just the
probability $F_M(p)$ from the percolation problem, which is
already known to have the appropriate limit. Note that $P_{(0;m)}(p_0)$ can also be described as the probability that on the MSF$(p_0)$ on the infinite BL there is a path to infinity passing through two specified vertices at separation $m$.

\begin{figure}
\includegraphics[width=3in]{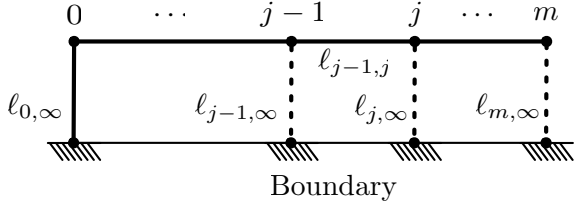}%
\caption{Construction of a path on the BL starting from
the origin (labeled $m$), passing through the vertex labeled $0$,
and connecting to the boundary, here shown as a shaded line. For
this path to lie on the MST, the edge costs must obey the set of
inequalities given in eq.\ (\ref{bethe_ineqs}). Note that each
vertical line (both solid and dashed) stands for a subtree, not
just a single edge, so this diagram depicts the entire BL. \label{figbethepath}}
\end{figure}

For later convenience, let us label the vertices so that $m$ is
the origin, and $0$ is the vertex at distance $m$ that the path on
the MST passes through. To begin with, consider $m>0$. For every vertex $j$ ($j=1$, \ldots, $m-1$) along the path from the origin to our chosen vertex, there is a set of paths on the BL connecting $j$ to the boundary through any of the $\sigma - 1$ side branches leaving the path from $0$ to $m$. We define
$\ell_{j,\infty}$ as the minimum over this set of the maximum edge
cost encountered along each of these paths (see Fig.\
\ref{figbethepath}). In the model with $\ell$ for each edge uniformly distributed in $[0,1]$, $\ell_{j,\infty}$ is the value of $p$ at
which a connection to infinity is first formed along any of these
$\sigma-1$ side branches from vertex $j$ in the bond percolation
process. Similarly, for $j=0$ ($j=m$), define $\ell_{0,\infty}$
($\ell_{m,\infty}$) as the minimum cost at which connection to
infinity is formed along any of the $\sigma$ branches
other than the path to $m$ ($0$) in the
percolation process. Then the desired probability can be written
as
\begin{multline}
\label{bethe_ineqs}
P_{(0,m)} (p) = \\
\mbox{Pr} \biggl[ \Bigl( \bigwedge_{j=1}^{m} \bigwedge_{i=0}^{j}
(\ell_{j,\infty} > \ell_{i-1,i}) \Bigr) \wedge\bigwedge_{j=0}^{m}
(\ell_{j-1,j} < p) \biggr] ,
\end{multline}
where $\wedge$ denotes a logical `and' and we simplify the
notation by defining $\ell_{-1,0} \equiv \ell_{0,\infty}$. Notice
that in forming the minimum cost tree connected to infinity, it is
immaterial whether $\ell_{i-1,i}>\ell_{j,\infty}$ for $i> j$,
provided that $\ell_{j',\infty}>\ell_{i-1,i}$ for $j'\geq i$. This
is the crucial fact in setting up a recurrence relation for
$P_{(0,m)}(p)$. Finally for $m=0$, we must make a special
definition. The most natural is that no direction for the path is
specified, and $P_{(0,0)}(p)$ is defined to be
$P_{(0,0)}(p)=P_\infty(p)=1-F(p)^{\sigma+1}$.

If the number of side branches from vertex $m$ ($m>0$) was the
same as those at vertices $j=1$, \ldots, $m-1$, namely $\sigma-1$,
the probability $P_{(0,m)}(p)$ would obey a recurrence relation.
Let us define $P'_{(0,m)}(p)$ to be defined the identical way with
this modification. (This corresponds to an alternative definition
of the BL that is sometimes used, in which the origin
alone in the tree has degree $\sigma$; it is used because it
simplifies the use of recurrence relations in a similar way as
here.) Then eq.\ \eqref{bethe_ineqs} can be explicitly constructed
as a recurrence relation, most easily by working with the derivatives %
\be%
\label{eq_phi_def} \Phi_j (p) = \frac{d}{dp}
P'_{(0;j)} (p). \ee %
The initial condition for the recurrence is %
\be%
\label{eq_1ptinit}
\Phi_0(p) = \frac{d}{dp} \bigl( 1-F(p)^{\sigma} \bigr). \ee %
The costs $\ell_{j,\infty}$ for $j=0$, \ldots, $m$, and
$\ell_{j-1,j}$ for $j=1$, \ldots, $m$ are all statistically
independent (note that $\ell_{m,\infty}$ now refers to a
collection of $\sigma-1$ branches, like the others). The
probability that $\ell_{j,\infty} < p$ is $1-F^{\sigma-1}(p)$ for
$j=1$, \ldots, $m$. The probability that $\ell_{j-1,j}<p$ is $p$.
The recurrence is then given by%
\bea%
\label{bethestep}%
\Phi_j(p)& = & F(p)^{\sigma-1} \left( p \Phi_{j-1}(p) + \int_0^p
\! dp' \,
\Phi_{j-1}(p') \right)\nonumber\\
&=&F(p)^{\sigma-1}\frac{d}{dp}\left( p \int_0^p \! dp' \,
\Phi_{j-1}(p') \right). \eea %
The factor of $F^{\sigma-1} (p)$ is the probability that the
$\sigma-1$ side branches at vertex $j$ are not connected to
infinity by $p$. In the bracket in the first line of eq.\
\eqref{bethestep}, the two terms correspond to the two cases that
the most expensive edge connecting the vertex $j$ to infinity on
the path, which has cost $p$, either is not or is the edge
connecting vertices $j-1$ and $j$, respectively. In the first case
the latter edge is already occupied (with probability $p$), and
connection occurs at $p$ somewhere in the remainder of path, with
probability density described by $\Phi_{j-1}(p)$, while in the
second the edge $j-1,j$ is the one that becomes occupied at $p$,
and the probability must be multiplied by the probability
$P'_{(0,j-1)}(p)$ that the rest of the path is already formed. We
write the recurrence in terms of a kernel $K$:
\bea%
\label{bethestep2a}
\Phi_{j} (p)&  =& \int_0^1 \! d p' \; K(p, p') \Phi_{j-1} (p'); \\
\label{bethestep2b} K (p, p')& =& F(p)^{\sigma-1}  \bigl[
\theta(p - p') +p \delta( p - p') \bigr] \nonumber \\
&=& F(p)^{\sigma-1}  \frac{d}{dp}\bigl[
p \theta(p - p')\bigr],\eea%
where $\theta(x)$ is the usual step function.

The kernel $K(p,p')$ in the recurrence has the meaning of a
conditional probability density. It is the probability density (in
the $p$ variable) that $j$ is first connected to infinity along
the specified path passing through $j-1$ at value $p$ (and is not
connected to infinity along any side branches), given that $j-1$
is first connected to infinity in the specified manner at value
$p'$. It is clear that this must vanish if $p<p'$. We note that if
the factors $F^{\sigma-1}$ are omitted, then we obtain the
corresponding conditional probability relevant to percolation, in
which connections to infinity along side branches are allowed,
instead of that relevant to MSTs. Similar interpretations apply to
the iterates of $K$ that we consider next.

The recurrence relation \eqref{bethestep}, and its
generalizations, allow us to compute correlation functions on the
BL. Iteration of the
recurrence requires iterated integrals of $K$, %
\begin{multline}
K^{\ast m} (p_1,p_{m+1}) \\
\begin{aligned}
&\equiv K\ast K\cdots \ast K(p_1,p_{m+1}) \\
&= \int_0^1 \! dp_2 \cdots \int_0^1 \! dp_{m} \, K(p_1,p_2) \cdots K(p_{m},p_{m+1}).
\end{aligned}
\end{multline}
The explicit step and $\delta$- functions in each factor of $K$
guarantee that $p_1 \geq p_2 \geq \cdots \geq p_{m+1}$. From the
definition we then have%
\be%
P_{(0;m)} (p_0) = \int_0^{p_0} \! dp
\, dp' \,F(p) K^{\ast m} (p,p') \Phi_0(p'), \ee %
where the factor $F(p)$ in the final integration restores the
correct number of branches at the origin.

\begin{figure}
\includegraphics[width=3in]{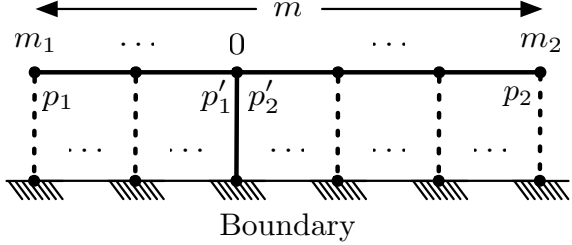}%
\caption{The set of paths contributing to $P_{(0;m_1,m_2)} (p_0)$,
drawn using the same conventions as figure \ref{figbethepath}.
For $C^{(2)} (m; p_0)$, we sum over all $m_1\geq0$, $m_2\geq0$ such that $m_1 +m_2 = m$. \label{figbethe2pt}}
\end{figure}

We can obtain multipoint correlation functions by introducing
branching into the path via additional chains of $K$s. For
example, the probability that two points at distances $m_1>0$,
$m_2>0$ from a vertex that we label $0$ are both connected to
infinity through $0$ by time $p=p_0$ is given by (see Fig.\ \ref{figbethe2pt})
\begin{widetext}
\be \label{eq_bethe_multi} P_{(0;m_1,m_2)} (p_0)= \int_0^{p_0} \!
dp_1 \, dp_2 \, dp'_1 \, dp'_2 \,F(p_1)F(p_2) K^{\ast
m_1}(p_1,p'_1) K^{\ast m_2}(p_2,p'_2) \Phi_{(0;0,0)} (p'_1,
p'_2), \ee
\end{widetext}
The initial distribution is
\be
\Phi_{(0;0,0)} (p,p') = \delta(p-p') \frac{d}{dp} (1- F^{\sigma-1}(p)),
\ee
because the
connection of $0$ to infinity must not use either of the two side
branches leading to $m_1$ and $m_2$. We then define the two-point
correlation function $C^{(2)}(m,p_0)$ as the probability that two
vertices at separation $m$ are connected to each other and to
infinity by $p_0$. For $m>0$, it is given by%
\be \label{eq_bethe_c2} %
C^{(2)} (m; p_0) = \sum_{m'=1}^{m-1}
P_{(0;m',m-m')} (p_0)+2P_{(0;m)} (p_0) . \ee %
The term $2P_{(0;m)} (p_0) $ covers the cases where the path to
infinity on the tree from one vertex passes through the other. For
$m=0$, we define $C^{(2)} (0; p_0)=P_{(0;0)}(p_0)=P_\infty(p_0)$
This construction of the two-point correlation function can be
immediately extended to multipoint correlations, giving the
probability that, by time $p_0$, several specified vertices are on
the same component of MSF$(p_0)$ and connected to infinity. These
will be used in section \ref{s_mft_treelevel} to analyze the geometry of the trees.

%%%%%%%%%%%%%%%%%%%%%%%%%%%%%%%%%%%%%%%%%%%%%%%%%

\subsection{Analysis of the iterated kernel}

To make further progress, we need to analyze the behavior of the
iterated kernel $K^{\ast m}$. First, for $P_{(0,m)}(p_0)$ there is
a simple result: substituting \eqref{eq_1ptinit} into
\eqref{bethestep} and using \eqref{eq_fdef} shows that $\Phi_1(p)
= \Phi_0(p)/\sigma$. That is, $\Phi_0$ is an right eigenfunction
of $K$ with eigenvalue $1/\sigma$. Then $\Phi_j(p) =
\Phi_0(p)/\sigma^j$
for $j=1$, \ldots, $m-1$. Then finally we obtain%
\be%
P_{(0,m)}(p_0)=\frac{\left[1-F(p_0)^{\sigma+1}\right]}
{(\sigma+1)\sigma^{m-1}}
=\frac{P_\infty(p_0)}{(\sigma+1)\sigma^{m-1}}.\ee%
In particular, if we put $p_0=1$ so that all edges have been
tested, then $P_{(0,m)}(1)=\sigma^{-(m-1)}/(\sigma+1)$. The
meaning of these results should be clear: due to the isotropy of
the BL, the path from the origin to infinity [which
exists at $p_0$ with probability $P_\infty(p_0)$] must pass
through one of the $(\sigma+1)\sigma^{m-1}$ vertices a distance
$m$ away, and all are equally probable. Thus the path is an
isotropic random walk on the BL (with no backtracking).
In section \ref{s_fractal_dim}, we will interpret this result in Euclidean space as a random walk with fractal dimension $D_p=2$.

All the right eigenfunctions of $K$ can also be obtained. We will require them to be integrable, in particular at $p=0$. The
eigenvalue equation for $K$, $\int_0^1
dp_2\,K(p_1,p_2)v_\lambda(p_2)=\lambda v_\lambda(p_1)$, becomes a
differential equation for $V_\lambda(p)=\int_0^p
dp'\,v_\lambda(p')$,
\be%
(\lambda-pF(p)^{\sigma-1})\frac{d}{dp}V_\lambda(p)
=F(p)^{\sigma-1}V_\lambda(p).\label {diffeq}\ee%
Locally in $p$, the general solution of the differential equation (\ref{diffeq}) is%
\be%
V_\lambda(p)=C\exp \int^p
dp'\,\frac{F(p')^{\sigma-1}}{\lambda-p'F(p')^{\sigma-1}}\ee %
($C$ is a constant). From the definition, $V_\lambda$ must obey the initial condition $V_\lambda(0)=0$, and be continuous at all $p$. Excluding the solution $V_\lambda(p)$ identically zero on $[0,1]$, the continuity of $V_\lambda(p)$ implies that $V_\lambda(0)=0$ is only possible if $\lambda$ obeys $\lambda=pF(p)^{\sigma-1}$ for some $p=p_\lambda\geq p_c$, and $V_\lambda(p)=0$ for $p<p_\lambda$
(the solution for $p_\lambda\geq p_c$ to $\lambda=pF(p)^{\sigma-1}$is unique). This means that
$\lambda$ lies in the interval $(0,1/\sigma]$. As $p_\lambda$ is approached from above, these solutions have a power law behavior, %
\be%
V_\lambda(p)\sim (p-p_\lambda)^{\alpha_\lambda+1}\ee %
as $p\to p_\lambda$ from above. Using the behavior of $F(p)$ near
$p_c$, we find $\alpha_\lambda\sim -2\sigma(p_\lambda-p_c)$ as
$p_\lambda\to p_c$ from above. Also, $\alpha_\lambda\to -1$ as
$p_\lambda\to 1$ ($\lambda\to0$), and $\alpha_\lambda$ always lies
between $0$ and $-1$. Thus the eigenfunctions $v_\lambda$ are
non-negative and integrable for all $\lambda\in(0,1/\sigma]$. For
$\lambda=1/\sigma$, $p_\lambda=p_c=1/\sigma$, and
$v_{1/\sigma}(p)\propto \Phi_0(p)$.

To make progress with higher-order correlation functions such as
\eqref{eq_bethe_multi}, we first introduce the generating function
(a discrete Fourier-Laplace transform), for complex $w$,
\be%
\label{bethe_laplace}
\tPhi_w(p) = \sum_{j=0}^\infty w^j \Phi_j(p).
\ee%
Inserting this into \eqref{bethestep2a} and summing over $j$ yields
\be%
\label{bethe_ker} \int_0^1 \! dp' \; \Bigl[ \delta(p-p') - w
K(p,p') \Bigr]\tPhi_w (p') = \Phi_0(p).
\ee%
In order to solve this integral equation, we find the resolvent
operator (or Green function) $g_w$ such that%
\be%
\label{eq_g_def} %
\int_0^1 \! dp_2
\Bigl[ \delta (p_1 - p_2) - w K(p_1,p_2) \Bigr] g_w (p_2, p_3) =
\delta(p_1-p_3). \ee %
The formal solution of \eqref{eq_g_def} is
simply %
\be %
\label{eq_g_sum} %
g_w  (p_1,p_2) = \delta(p_1-p_2)+\sum_{j=1}^\infty w^j K^{\ast
j}(p_1,p_2). \ee%
Because the set of eigenvalues of $K$ is the interval
$(0,1/\sigma]$, the series cannot be expected to converge if
$|w|>\sigma$. This will be directly confirmed later. From the
structure of $K$, we expect in general that $g_w$ is zero for
$p_1<p_2$. Because of the $\delta$-function term in $K$, $g_w$
contains a $\delta$-function term as well as a smooth piece. For
$|w|<\sigma$, the $\delta$-function part is clearly
$g_w(p_1,p_2)=(1-wp_1F(p_1)^{\sigma-1})^{-1}\delta(p_1-p_2)+\ldots$,
where the omitted parts are ordinary functions. For $w$ real and
in the interval $[\sigma,\infty)$,  the coefficient of the
$\delta$-function blows up at $p_1=p_2=1/w$ and at
$p_1=p_2=p_{1/w}$, in terms of the values $p_\lambda$ related to
the eigenvalues $\lambda$ of $K$ as above.

Equation \eqref{eq_g_def} can be converted into a differential
equation by defining $G_w(p_2,p_3)=\int_0^{p_2}
dp'_2\,g_w(p'_2,p_3)$; then
\begin{widetext}
\be \label{eq_big_g_def} \bigl(1-w p_1 F^{\sigma-1} (p_1) \bigr)
\frac{d}{dp_1} G_w (p_1,p_2) -wF^{\sigma-1}(p_1) G_w(p_1,p_2)=
\delta(p_1-p_2). \ee
\end{widetext}%
For $p_1 \leq \pc$ we have $F(p_1) =1$, and  eq.\
\eqref{eq_big_g_def} is solved [using the initial condition
$G_w(0,p_2)=0$] by%
\be %
\label{subcrit_pgator}%
G_w(p_1,p_2) = \frac{\theta(p_1-p_2)}{1-w p_1} , \ee%
which vanishes if $p_2>p_c$.

The full solution to \eqref{eq_big_g_def}, valid for all $p_1$ and
$p_2$, is obtained by solving the equation without the
$\delta$-function term for $p_1>p_2$ as in eq.\ (\ref{diffeq})
above (with $\lambda=1/w$), and choosing the constant to obtain
the correct discontinuous behavior at $p_1=p_2$ (because of the
$\delta$-function). Because of the initial condition
$G_w(0,p_2)=0$, we impose the requirement that $G_w$ must vanish
for $p_1<p_2$ for all $p_2$. The result is%
\begin{multline}
\label{supercrit_pgator}
G_w (p_1,p_2) = \frac{\theta(p_1-p_2)}{1-w p_2 F^{\sigma-1}(p_2) }\\
\times \exp \int_{p_2}^{p_1} dp'\,\frac{wF(p')^{\sigma-1}}{1-wp'F(p')^{\sigma-1}}.
\end{multline}
Then $g_w$ is obtained as $g_w(p_1,p_2)=\partial
G_w(p_1,p_2)/\partial p_1$. Both $g_w$ and $G_w$ can be seen to be
complex analytic for $w\not\in [\sigma,\infty)$. $g_w$  exhibits
singular behavior if $w$ is real and in $[\sigma,\infty)$ (as
expected), and nowhere else, which means that the spectrum of $K$
is precisely the interval $(0,1/\sigma]$. For
$w\not\in[\sigma,\infty)$, there is no solution of the homogeneous
equation obeying the initial condition that could be added to the
solution. These results agree with the earlier statements for the two  regimes $|w|<\sigma$ and $p_1<p_c$.

To extract the large $m$ behavior of $K^{\ast m}$, we write%
\be%
K^{\ast m} (p_1,p_2) = \frac{1}{2 \pi i} \oint \! dw \; \frac{ g_w
(p_1,p_2)}{w^{m+1}}, \ee %
where the contour is a small circle around the origin. Because
$g_w$ is analytic in $w$ except on $[\sigma,\infty)$, we can
increase the radius of the contour until we hit the first
singularity in $w$, which is located on the real axis at the value
$w_c$ defined by $1-w_c p_2 F^{\sigma-1}(p_2) =0$. Parameterizing the contour as  $w = w_c e^{-i \tau} $, the most strongly divergent piece of the integrand at this singularity will
determine the large-$m$ behavior of $K^{\ast m}$ via
\be
\label{eq_fourier_power}
\begin{aligned}
\int_{-\pi}^\pi \! d \tau \; \frac{e^{i \tau m}}{(i \tau)^z} &= \frac{2 \pi m^{z-1}}{\Gamma(z)} \\
& \quad +i m^{z-1} e^{-i \pi z} \Gamma(1-z; -i m \pi) \\
& \quad - i m^{z-1}e^{i\pi z} \Gamma(1-z; i m \pi),
\end{aligned}
\ee
where $\Gamma(z; t_0) = \int_{t_0}^\infty \! dt \; e^{-t} t^{z-t}$
is the incomplete gamma function. The first term on the right-hand
side of \eqref{eq_fourier_power} comes from the part of the contour
coming in from $\tau = 0^+ +i \infty$, encircling the singularity,
and returning to $\tau = 0^- + i \infty$. If $z$ is not an integer,
there will be a branch cut from $\tau = 0$ to $\tau = + i \infty$,
 but the same answer is obtained regardless. The last terms come
 from the segments of the contour running from $\tau = -\infty$
 to $\tau = -\pi$ and $\tau = \pi$ to $\tau = \infty$; using the
 asymptotic form of the incomplete gamma function we see that
\begin{multline}
i m^{z-1} \left( e^{-i \pi z} \Gamma(1-z; -i m \pi) -e^{i\pi z}
\Gamma(1-z; i m \pi) \right) \sim \\
\frac{2}{m \pi^z} \sin \left( \frac{2 \pi m + \pi (1-z)}{2} \right)
 + \bigO(1/m^2),
\end{multline}
and so these terms may be neglected relative to the first as $m\to \infty$.

In order to make progress in closed form we specialize to $ \sigma
=2$. Using \eqref{eq_fdef}, $F(p) = (1-p)/p$ for $p>p_c = 1/2$ and%
\begin{multline} %
\exp\int_{p_2}^{p_1}
dp'\,\frac{wF(p')^{\sigma-1}}{1-wp'F(p')^{\sigma-1}} \\
= \left(\frac{p_2^w (1-w(1-p_1))}{p_1^w (1-w(1-p_2))} \right)^{\frac{1}{w-1}},
\end{multline} %
taking $p_1 > p_2 > p_c$. Defining
$h_w(p) = 1-w(1-p)$ for brevity, %
\be %
\label{eq_sig2_g} %
g_w
(p_1,p_2) = \frac{d}{dp_1} \left[ \frac{\theta(p_1-p_2)}{h_w(p_2)}
\left( \frac{p_2^w h_w(p_1)}{p_1^w h_w(p_2) }
\right)^{\frac{1}{w-1}} \right]. \ee %
It might appear that
\eqref{eq_sig2_g} has an essential singularity at $w = 1$, but in
fact %
\be %
\lim_{w \to 1} g_w (p_1,p_2) = \frac{d}{dp_1}
\left[\frac{\theta(p_1-p_2)}{p_1} e^{\frac{1}{p_1} -
\frac{1}{p_2}} \right], \ee %
and the singularity with smallest
$|w|$ is located at $w_c = 1/(1-p_2)$; $g_w$ may be rewritten as%
\be %
g_w (p_1,p_2) = \frac{d}{dp_1} \frac{ \theta(p_1-p_2)}{ (w_c
-w)^{\frac{w}{w-1}}} \left( \frac{p_2^w h_w(p_1)}{p_1^w (1-p_2)^w}
\right)^{\frac{1}{w-1}} . \ee %
Because $1/2 \leq p_2 \leq 1$, we
know $1 \leq w_c/(w_c -1) \leq 2$ and the factor of $(w_c-w)$ is
responsible for all divergences as $w \to w_c$. To find the
leading-order divergence of this factor, we use
\begin{widetext}
\be
(w_c -w )^{-\frac{w}{w-1}} = (w_c -w )^{-\frac{w_c}{w_c-1}} + \bigO \left( (w_c -w )^{-\frac{w_c}
{w_c-1}+1}
\log (w_c -w ) \right),
\ee
which, with $w = w_c e^{-i \tau} $, diverges as $(i \tau)^{-1/ p_2}$.
We may
now use \eqref{eq_fourier_power} to evaluate
\be%
K^{\ast m} (p_1,p_2)= \frac{1}{2 \pi w_c^m} \int_{-\pi}^\pi \!
d\tau \; e^{i \tau m} g_{w_c e^{- i \tau}}(p_1,p_2).
\ee%
Because of cancellations that take place in \eqref{eq_sig2_g} when
$p_1 = p_2$,
we must take the derivative with respect to $p_1$ explicitly before
the $m \to
\infty$ limit. The result is
\be%
\label{eq_big_km}
K^{\ast m} (p_1,p_2) \sim \delta (p_1-p_2) (1-p_2)^m +
\theta(p_1-p_2) \frac{(1-p_1)(1-p_2)}{p_1(p_1-p_2)^2} \left(
\frac{p_2(p_1-p_2)}{p_1(1-p_2)} \right)^{1/p_2} \frac{(1-p_2)^m \,
m^{1/p_2 -1}}{\Gamma(1/p_2)}.
\ee%
\end{widetext}

%%%%%%%%%%%%%%%%%%%%%%%%%%%%%%%%%%%%%%%%

\subsection{Asymptotics of the two-point function and mass}
\label{sec_asymp}

In this section we complete the calculation of the important
correlation function $C^{(2)}$. We present two versions: an
essentially exact version for $\sigma=2$, and an asymptotic
calculation valid for all $\sigma$ in the region $p_0-p_c$ small.

If we make no further approximations, when we calculate
correlation functions such as \eqref{eq_bethe_c2} it is clearly
easier to integrate over $p_2$ first and then find the large-$m$
behavior rather than attempt to integrate \eqref{eq_big_km}
directly over $p_2$. As an example we now calculate the asymptotic
behavior of $C^{(2)} (m; p_0)$. Inserting $\delta_{m, m_1 + m_2}$
in \eqref{eq_bethe_c2} gives
\begin{widetext}
\be %
\label{eq_big_c2} C^{(2)} (m; p_0) = \int_0^{p_0} \! dp_1 dp_2
\;  \frac{1}{2 \pi w_c^m} \int_{-\pi}^\pi \! d\tau \; e^{i \tau m}
\tPhi_{w_c e^{-i \tau},w_c e^{-i \tau}} (p_1, p_2), \ee%
where %
\be %
\label{eq_big_phiww} %
\tPhi_{w_1,w_2} (p_1, p_2) =
\int_0^1\! dp'_1 dp'_2 \; g_{w_1} (p_1, p'_1) g_{w_2} (p_2, p'_2)
\Phi_{(0;0,0)}(p'_1,p'_2). \ee %
with $\Phi_{(0;0,0)}(p'_1, p'_2) =
\theta(p'_1-p_c) \delta(p'_1-p'_2) / {p'}_1^2$ for $\sigma=2$.
This integral is done easily: %
\be %
\tPhi_{w,w}(p_1, p_2) =
\frac{d}{dp_1} \frac{d}{dp_2} \frac{1}{w+1} \left( \frac{h_w(p_1)
h_w(p_2)}{p_1^w p_2^w} \right)^{\frac{1}{w-1}} \left[ \left(
\frac{1}{2-w} \right)^{\frac{w+1}{w-1}} - \left( \frac{p}{h_w(p)}
\right)^{\frac{w+1}{w-1}} \right], \ee %
where $p = \min(p_1, p_2)$. Clearly, since the integrand in
\eqref{eq_big_phiww} has no support below $p_c$, we obtain a
branch cut starting at $w_c = 1/p_c = 2$ and this is the maximum
radius the contour in $w$ may take. Since the only divergence as
$w \to 2$ comes from the first term in the square brackets, we
obtain %
\be %
\tPhi_{2 e^{-i \tau},2 e^{-i \tau}}( p_1, p_2) =
\frac{d}{dp_1} \frac{d}{dp_2} \frac{h_2(p_1) h_2(p_2)}{p_1^2
p_2^2} \left[ \frac{1}{3} \left( \frac{1}{2i \tau} \right)^3 +
\bigO\left( \frac{ \log \tau}{\tau^2} \right) \right]. \ee%
\end{widetext}
Then, using
\be%
\int_{-\pi}^\pi \! d \tau \; \frac{e^{i \tau m}}{(i \tau)^{3}} \sim m^2
\pi + \bigO(1/m)
\ee%
and
\be%
\int_{p_c}^{p_0} \! dp \; \frac{d}{dp} \frac{h_2(p)}{p^2} =
\frac{2 p_0 -1}{p_0^2}
\ee%
we finally obtain
\begin{multline}
\label{eq_c2_asymp}
C^{(2)} (m; p_0) \sim \\
\left( \frac{2 p_0 -1}{p_0^2} \right)^2 2^{-m} \left(
\frac{m^2}{48} + \bigO( m \log m) \right).
\end{multline}
We see that $C^{(2)} (m; p_0) $ scales as $m^2/\sigma^m$ for large
$m$, and this behavior is obtained for \emph{any} $p_0
p_c$.

%referee
%clarified approximation used below

%Next we present an alternative calculation valid for all $\sigma$
%when $p_0-p_c$ is small. We will define $\delta p = p-p_c$ (but e.g.\
%$\delta(p_1-p_2)$ is a $\delta$-function as usual). First we write%
%\bea%
%pF(p)^{\sigma-1}&\simeq &p_c-\delta p,\\%
%\label{eq_F_approx}
%F(p)^{\sigma-1}&\simeq& 1-2\sigma\delta p,\eea%
%valid as $p\to p_c$ from above. Then, using \eqref{supercrit_pgator}, $G_w$ can be calculated for
%$w\not\in[\sigma,\infty)$,%

Next we present an alternative calculation valid for all $\sigma$
when $p_0-p_c$ is small. We will define $\delta p = p-p_c$ (but e.g.\
$\delta(p_1-p_2)$ is a $\delta$-function as usual). We then have
\be
\label{eq_F_approx}
pF(p)^{\sigma-1} \simeq p_c-\delta p,\
\ee
valid as $p\to p_c$ from above. Then, using \eqref{supercrit_pgator}, $G_w$ can be calculated for $w\not\in[\sigma,\infty)$. Because all the singular behavior as $w \to \sigma$ arises from the denominator of the integrand of \eqref{supercrit_pgator}, we may approximate $F^{\sigma-1}(p) \simeq 1$ in the numerator and
\be%
G_w(p_1,p_2)=\theta(p_1-p_2)\frac{\delta p_1+1/w-p_c}{w(\delta
p_2+1/w-p_c)^2}\ee%
for $\delta p_1$, $\delta p_2 $ both small and positive. Then if
we consider the transform %
\be%
\widetilde{C}^{(2)}_w(p_0)=\sum_{m=0}^\infty w^m C^{(2)}(m,p_0)\ee%
we notice that the transform of the sum over $m_1$ in eq.\
(\ref{eq_bethe_c2}) becomes simply a product of $g_w$s inside the
integral in eq.\ (\ref{eq_bethe_multi}), and we neglect the
factors $F(p_1)$, $F(p_2)$ as these do not affect the leading $m$
dependence. Further we can neglect $2P_{(0;m)}$ as it falls off
faster than the term we keep. Also $\Phi_{(0;0,0)}(p_1,p_2)=
2\sigma\delta(p_1-p_2)$ for $p_1$ above $p_c$, and zero below.
Finally, we will estimate the expected mass
inside radius $m$, %
\be%
\overline{M(m,p_0)}=C^{(2)}(0;p_0)+ \sum_{m'=1}^m (\sigma+1)\sigma^{m'-1}
C^{(2)}(m',p_0)
\label{Mdeff}\ee%
directly, as this is similar to the definition of the transform of
$C^{(2)}$: we simply evaluate the transform at $w=\sigma(1-1/m)$,
which cuts off the sum at around $m$. This is a value at which
$G_w$ is not singular. Thus we have to calculate%
\begin{multline}
\widetilde{C}^{(2)}_w(p_0)= 2\sigma(\delta p_0+ 1/w-p_c)^2\\
\times \int_{0}^{\delta p_0} d\delta p'\,\frac{1}{w^2(\delta p'+ 1/w-p_c)^4}.
\end{multline}%
For fixed $p_0$, the dominant contribution comes from the lower
limit, and contains $(1-wp_c)^{-3}$ times factors that go to
constants as $w\to\sigma=1/p_c$. Hence the mass behaves as %
\be%
\overline{M_{\rm soft}(m,p_0)}\sim \frac{2}{3}\sigma(\sigma+1)\delta p_0^2 m^3\label{massform}\ee%
as $m\to\infty$ [we inserted a factor $(\sigma+1)/\sigma$ to
account for the number of neighbors $\sigma+1$ at the first step
in $\overline{M(m,p_0)}$, as in eq.\ (\ref{Mdeff})]. This method in fact
differs from the definition above in using a soft cut-off for the
sum over $m'$ instead of a hard one, $m'\leq m$. Now that the form
of the summand is known, we can evaluate it using either form of
cutoff. Hence we find that for the {\em hard} cut-off, the result is
smaller by a factor $6$:
\be%
\overline{M(m,p_0)}\sim \frac{1}{9}\sigma(\sigma+1)\delta p_0^2 m^3.\label{massformhard}\ee%
Thus the correlation function behaves as%
\be%
C^{(2)}(m,p_0)\sim \frac{1}{3}(\sigma \delta p_0)^2
m^2\sigma^{-m},\label{corrfnform}\ee%
which agrees with the result for $\sigma=2$.

Equations (\ref{massformhard}) and (\ref{corrfnform}) are the main results of this section. From the structure of
the expression for $G_w$, the lower limit always dominates, so
this $m$ dependence holds for all $p_0>p_c$, at sufficiently large
$m$. More precisely, the results are valid only if $\delta p_0$
is greater than order $1/m$. This means that $m$ is much larger
than the correlation length at $p_0$, which is proportional to $1/(\delta p_0)$. The main contribution to the integral is from $\delta p'$ less than of order $1/m$. This is in agreement with the
``superhighways'' idea \cite{wu_superhwy_1,wu_superhwy_2,wu_rrns}.

Notice also that the factor $\delta p_0^2$ is present in both results because the probability that one vertex is connected to infinity is $P_\infty(p_0)\simeq 2\sigma(\sigma+1)\delta p_0/(\sigma-1)$, for $\delta p_0$ small and positive, and so is $\propto\delta p_0^2$ for two vertices (the two events are uncorrelated, because the two points are separated by more than the correlation length). The dependence on
$\sigma$ is the same within subleading terms of relative order
$1/\sigma$.

\subsection{General $k$th-order correlation functions and moments of the mass}
\label{s_mft_treelevel}

We may extend the method to estimate asymptotics of correlation functions of any order $k$, that is the probability $C(i_1,\ldots,i_k;p_0)$ that some given set of vertices (labeled $i_1$, $i_2$, \ldots, $i_k$) are on the same connected component of MSF$(p_0)$ and that this component is infinite. The procedure is straightforward: to compute the $k$-point correlation function for a given set of $k$ distinct vertices, one draws the smallest subtree of the BL such that all $k$ given vertices are connected, and choose any vertex on this subtree as the root point, along which the connection to infinity occurs in the MSF$(p_0)$ (eventually, we will sum over the possible root points). Thus the leaves of the subtree (i.e.\ the degree 1 vertices) must be among the given $k$ vertices, but if any of the $k$ given vertices are not leaves, they can be anywhere on the subtree. Starting from the root, we propagate out to (or possibly through) each of the $k$ given vertices, along the subtree. The subtree can be viewed as made of chains of edges connected by degree-2 vertices, with the ends of the chains at either (i) the root point, which has degree $\geq 1$, (ii) the leaves of the subtree, or (iii) vertices of degree $>2$ other than the root point. For each such chain $e$ of $m_e$ steps, we associate the iterated kernel $K^{\ast m_e}(p_i,p_j)$, where the labels $i$, $j$ are associated to the two ends of the chain, with $i$ the end further from the root point. For the initial distribution at the root, if there are $n$ chains leaving it ($n\leq \sigma$), we generalize $\Phi_{(0;0,0)}$ to
\begin{multline}
\Phi_{(0;0, \ldots, 0)}(p_1, \ldots, p_n) =\\
 \left[\frac{d}{dp_1} \left( 1- F^{\sigma+1-n}(p_1) \right)\right] \prod_{j=2}^n \delta(p_1 - p_j) .
\end{multline}
Similarly, by comparing $dP_\infty (p_0)/dp_0 $ with $\Phi_0(p')$ and $\Phi_{(0;0,0)}(p'_1, p'_2) $, we see that at a vertex of the subtree of degree $n\neq 2$, we must associate with it a factor
\be
v_n (p_1, p_2 \ldots, p_n) = F^{2-n}(p_1)
\prod_{j=2}^n \delta(p_1 - p_j).
\ee
After multiplying together all these factors, we must integrate over all the parameters like $p_i$ between the limits $0$ and $p_0$. There are two of these parameters for each chain on the subtree; clearly some could be eliminated using the $\delta$ functions. Finally, we must sum over all possible root points on the tree. This procedure yields $C(i_1,\ldots,i_k;p_0)$ (unlike our earlier description for the $k=2$ case, there are no exceptions to this prescription for cases of vertices coinciding with each other or with the root point).

The higher-point correlation functions can be used to calculate higher moments of the mass, $\overline{M(m,p_0)^k}$. These are the average of the $k$th power of the sum over positions at distance less than $m$ from the origin of the ``indicator function'' that is one if and only if the vertex is on the same connected component of MSF$(p_0)$ as the origin. $\overline{M(m,p_0)^k}$ is equal to the sum of the $k+1$-point correlation function $C(i_1,\ldots,i_{k+1};p_0)$ over all positions of $i_2$, \ldots, $i_{k+1}$ within $m$ steps of the origin at $i_1$.

For $m$ large, the largest contribution to $\overline{M(m,p_0)^k}$ will come from configurations of $i_l$, ($l=1$, \ldots, $k+1$) for which, in the subtree in the calculation of $C(i_1,\ldots,i_{k+1};p_0)$, all the given vertices are at its leaves, the root point has degree $2$, and the vertices of degree $>2$ have degree $3$. For these there are $2k$ chains (iterated kernels) in the subtree. We can estimate the power of $m$ as $m\to\infty$ using the same approximations as for $k=1$. The sums over position are estimated by using the propagator $g_w$ in place of all $K^{\ast m_e}$'s, with $w=\sigma(1-1/m)$ in each one. The factors at the vertices of the subtree (other than $\delta$-functions) can be dropped, at least when $\delta p_0$ is small (and for larger $\delta p_0$ do not affect the scaling behavior). The integrals over the $p_i$ associated with the leaves can be done by using $G_w$ in place of $g_w$ for these chains. The remaining integrals over $p_i$'s associated with the other vertices and the root point are dominated by the lower limit $\delta p_i=0$, and can be estimated by power counting. As each additional leaf on the subtree leads to an extra factor $G_wg_w$ and one additional integral similar to that at the root (as for $k=1$ above), this yields finally (neglecting constant factors)%
\be%
\overline{M(m,p_0)^k}\sim m^{3k}\sim [\overline{M(m,p_0)}]^k.\ee
As all $k$th moments scale like the $k$th power of the first moment, this means that the (random) mass $M(m,p_0)$ does not have a very broad distribution, and its typical behavior is well-described by its expected value. Hence the connected components of MSF$(p_0)$ on the BL are not multifractals.

\subsection{Fractal dimensions}
\label{s_fractal_dim}

So far we have developed a method for computing correlation
functions on the BL, while our real interest is in
lattices in Euclidean space of dimension $d$. For sufficiently
high $d$, we would expect that a mean field theory holds for
quantities such as exponents; this assertion will have to be
justified {\it post hoc} in a subsequent paper, by a perturbation analysis of corrections due to fluctuations neglected in the mean field theory. The BL results provide the mean-field theory results, once we
have explained how to convert them to apply to Euclidean space.

For a hypercubic lattice on Euclidean space, if we choose a path (starting from the origin) randomly (with equal probability for each), then it behaves as a random walk, and after $m$ steps will be of order $\sqrt{m}$ in Euclidean distance from the origin. In the set of all paths from the origin, any two paths initially coincide but ultimately part company. If we neglect the possibility that they subsequently intersect (and also that a path may intersect itself, including by backtracking), then the union of the paths forms a tree, equivalent to the Bethe lattice with $z=2d$. Hence in this correspondence, separations on the BL behave like distances {\em squared} on the Euclidean lattice \cite{fisher_essam}, %
\be%
\label{eq_bl_to_euclid}
m\sim r^2\ee%
which allows us to infer mean-field scaling dimensions from the BL theory. More formally, to apply the Bethe lattice results as an approximation for the lattice, in equation \eqref{bethestep} we must now sum over all neighbors of the given site, since the path may go through the site in any direction. Equation \eqref{bethe_laplace} needs to be
replaced by a Fourier transform \be \tPhi_{\k} (p) =
\sum_{\bfx} e^{i \k \cdot \bfx} \Phi_{\bfx} (p), \ee which means
that in \eqref{bethe_ker} and all subsequent equations we make the
substitution
\be%
\label{c_wavevector} w \to \sum_{\hat{n}_j} e^{-i \k \cdot
\hat{n}_j} = 2d - \k^2 + O(| \k|^4),
\ee%
where $\{ \hat{n}_j \}$ are the basis vectors of the hypercubic
lattice. This leads to the same relation \eqref{eq_bl_to_euclid}. 

%referee
%added clarification, extra references

%Note that we may also derive the relation \eqref{eq_bl_to_euclid} from %an explicit embedding of the BL into infinite-dimensional Euclidean %space, constructed as follows \cite{bl_to_euclid}. Assuming $\sigma$ %even, we may label all vertices $m$ steps from the origin of the BL by %strings of the form $\lambda_1 \lambda_2 \cdots \lambda_m$, where %$\lambda \in \{ 1 \cdots \sigma \}$ labels the outgoing edge chosen to %reach the vertex in question. The BL may be embedded in a hypercubic %lattice of formally infinite dimension by mapping $\lambda$ to a string %$x \in \mathbb{Z}^{\sigma/2}$, with $x_{\lceil \lambda/2 \rceil} = 1$ %for $\lambda$ odd, $-1$ for $\lambda$ even, and $x_i = 0$ for $i \neq %\lceil \lambda/2 \rceil$. Then the vertex $\lambda_1 \lambda_2 \cdots %\lambda_m$ is mapped to the point with Euclidean coordinates $\vec{r} = %(x_{\lambda_1}x_{\lambda_2} \cdots x_{\lambda_m} 0 \cdots )$. The %Euclidean distance from the origin to this point is
%\be
%\sqrt{|\vec{r}|^2} = \sqrt{m}.
%\ee
%This relationship has also been employed in several mathematically %rigorous treatments of percolation \cite{math_perc}.

%end addition

We established that the probability the path from any vertex
to infinity passes through a given vertex a distance $m$ away behaves
as $P_{(0;m)}(p_0)\sim \sigma^{-m}$; summing this over all sites
within a ball of radius $m$ on the BL gives %
\be %
P_{(0;0)}(p_0)+\sum_{m'
= 1}^m (\sigma +1)\sigma^{m'-1} P_{(0;m')} (p_0) \sim m. \ee %
Then we expect that in Euclidean space, the mass of the path lying
within radius $R$ scales as $M_p(R,p_0)\sim R^2$, consistent with the
picture of this path (mentioned earlier for the BL) as
a random walk, with dimension $D_p=2$. This is the same as the
dimension of the backbone of a critical percolation cluster for
$d\geq d_c=6$ \cite{stah}.

Similarly, the mass of the component connected to the origin on
the BL, $M(m,p_0)\sim m^3$ becomes $M(R,p_0)\sim R^6$ within
radius $R$ in Euclidean space, meaning that a connected component
of the MST has a mean-field fractal dimension $D=6$. Because the
union of the spanning trees fills the lattice, this strongly
suggests that the critical dimension of the MST will be $d_c=6$,
as discussed in Section \ref{s_ns_gs}. This is the same critical dimension as for percolation (at threshold). We emphasize again that the result is valid for $p_0$ greater than of order $1/m\sim 1/R^2$. Since the correlation length $\xi$ behaves as $|p_0-p_c|^{-\nu_\text{perc}}$, with $\nu_\text{perc}=1/2$ for $d>6$, this means it holds for $R>\xi$, and thus involves distance scales at which ordinary correlations for percolation decay exponentially.

\subsection{Poisson-weighted infinite tree}

Another mean-field model for MST (and other random optimization
problems), which mimics the continuum model in Euclidean space, is
based on the Poisson-weighted infinite tree (PWIT) \cite{ald1}.
This is a tree with infinite degree at each vertex, and the
weights (or costs) $\ell$ for the edges emanating from each vertex
are given by a Poisson process on $\ell\geq0$ with density
$\rho(\ell)\propto \ell^{d-1}$. This is the same as the measure on
the set of distances between the points of the uniform Poisson
process on $\mathbb R^d$. Thus the model is obtained by taking the
continuum model, and ignoring all correlations induced by the
geometry of space, keeping only the probability distribution for
separations of points.

The PWIT can be viewed as the limit $\sigma\to\infty$ of the iid
BL model with degree $\sigma+1$ we have used above,
with a different probability distribution on the edges (as we have
seen, the choice of this does not affect the geometry of the MST).
For $\sigma+1$ neighbors, we can cut off the distribution with
density $\rho(\ell)$ at $\ell={\cal O}((\sigma+1)^{1/d})$, then
study the finite-degree trees in the limit. For each $d\geq 1$ in
this model, the percolation threshold $\ell_c$ is non-zero, and
because the behavior is dominated by edges close to threshold, the
results for the PWIT will be in the same
universality class as the BL model above. Indeed, if we
set $\ell=(\sigma+1) p$, the BL model as
$\sigma\to\infty$ coincides with the $d=1$ PWIT. Notice that the
limit of our expressions exists, because by writing
$F(p)=1-G(\ell)/\sigma$, then $F^{\sigma-1}\to e^{-G}$, and the fixed-point equation for $F$ becomes
\cite{ald2}%
\be%
G(\ell)=\ell(1-e^{-G(\ell)}).\ee%
In the transforms, we should also set $s=w/\sigma$. Then the limit of our theory makes sense for the masses $M$, $M_p$ within $m$ steps of the origin: $\overline{M}\sim \frac{2}{3}\delta \ell^2 m^3$ (again, $\delta\ell=\ell-\ell_c$, where here $\ell_c=1$). For the PWIT one would naturally wish to express such quantities in terms of the distance $\ell$ defined as the sum of the $\ell_e$'s along a path on the tree, and since most edges accepted are either below, or not far above, $\ell_c$, these masses $M(\ell)$, $M_p(\ell)$ scale the same way, and again $\ell\sim R^2$ because the paths are random walks. When this measure of distance is used, the number of connected components inside $\ell$ is also well-behaved. Note that because it is believed that continuum percolation is in the same universality class as bond
percolation, we would expect universal properties of the continuum
MST to be the same as those of the lattice model anyway, so that
all of these conclusions are consistent.

%referee
%new section added
\subsection{Random, locally treelike graphs}

The results of the previous sections admit a simple generalization to certain classes of random graphs. We consider a tree where the coordination number of each vertex is an iid random variable distributed according to $\rho(\sigma)$.

In what follows, averages with respect to $\rho$ are denoted by angle brackets. The fixed-point equation for $F(p)$ becomes
\be
F(p) = 1-p + p \sum_{\sigma=0}^\infty \rho(\sigma) F(p)^\sigma.
\ee
Again, $F(p)$ is defined as the smallest solution at fixed $p$. In what follows we assume $1 < \langle \sigma \rangle < \infty$, which are the conditions necessary for the graph to admit a conventional percolation transition: $F(p) = 1$ for $p \leq p_c = 1/\langle \sigma \rangle$, $F(p) < 1$ for $p > p_c$, and we can construct a nontrivial ensemble of MSTs under wired boundary conditions at infinity.

At this point we find it helpful to multiply the kernel defined in \eqref{bethestep2b} by a factor of $\sigma$. This has the effect of summing the connectedness functions over lattice sites (done in, e.g., \eqref{Mdeff}) simultaneously with averaging over edge costs. This modification allows us to relate the spectrum and eigenfunctions of the random graph kernel $\langle \sigma K \rangle$ back to those obtained above: e.g, $\langle \Phi_0(p)\rangle$ is an eigenfunction of $\langle \sigma K \rangle$ but \emph{not} of $\langle K \rangle$.

The entire calculation goes through as before, with occurrences of $F(p)^{\sigma -1}$ in the BL Green's function replaced by $\langle \sigma F(p)^{\sigma-1} \rangle$ and the transform variable $w$ rescaled by $1/\langle \sigma \rangle$. Under the additional assumption $\langle \sigma^2 \rangle < \infty$, one may repeat the asymptotic analysis of section \ref{sec_asymp} and obtain
\be
\left\langle \overline{M_{\rm soft}(m,p_0)} \right\rangle \sim \frac{2}{3}\langle \sigma(\sigma+1) \rangle \delta p_0^2 m^3.
\label{massformavg}
\ee
The generalization extends to the higher moments of the cluster mass discussed in section \ref{s_mft_treelevel}.

One may also consider ``quenched'' moments of cluster masses, of the form
\be
\mathcal{M}_{k,\ell} (m,p_0) \equiv \left\langle \left( \overline{M(m,p_0)^k} \right)^\ell \right\rangle
\ee
for $\ell >1$. These quantities cannot readily be calculated with the techniques discussed above and are beyond the scope of this paper.

\section{Conclusion}
\label{s_conclude}

In this paper, we have achieved the following results. For a finite graph, we defined a process MSF$(p_0)$ which is a random forest that becomes the MST for $p_0=1$. Using the Bethe lattice (BL) with wired boundary conditions, and taking the infinite size limit, we showed that the infinite connected components of MSF$(p_0)$ on the BL contain of order $m^3$ vertices within $m$ steps of any vertex on this component, for any $p_0$ greater than the value $p_c$ of the threshold for bond percolation on the BL. This result is essentially rigorous. Transferring it (heuristically) to Euclidean space, this means that the mass of an infinite connected component of MSF$(p_0)$ within a ball of radius $R$ scales as $R^D$ with $D=6$, for $d$ sufficiently large and $p_0$ greater than the value $p_c$ of the threshold for bond percolation on the lattice used. This then implies that for $d>6$ there are of order $R^{d-6}$ large connected components that intersect such a ball. We also gave a non-rigorous second argument for these results, using scaling ideas (this argument directly addresses the critical dimension $d_c$ above which the results hold). The results also hold (rigorously) for the Poisson-weighted infinite tree, and (heuristically) for the continuum MST model in Euclidean space.

Following the reasoning of NS \cite{ns_1,ns_2}, these results for the MST imply that the strongly-disordered spin-glass model has an uncountable number of ground states for $d>6$, of which of order $2^{\bigO(R^{d-6})}$ can be distinguished within a ball of radius $R$. For $d\leq 6$, the logarithm of the number of ground states is smaller than any power of $R$, and possibly only of order one, or simply one (with probability one).

\begin{acknowledgments}
We would like to thank D. Stein and C. M. Newman for helpful discussions. This work was supported by NSF grant no. DMR--0706195.
\end{acknowledgments}

\bibliographystyle{apsrev}

\end{document}